\documentclass[fp,twocolumn]{jpsj3}
\usepackage{txfonts}
\usepackage{url}

\usepackage{xcolor}
\usepackage{bm}
\usepackage{algorithm}
\usepackage{algpseudocode}

\usepackage{amsmath}
\usepackage{amssymb}

\title{Geometric Characteristics of Subproblems in Ising-Machine-Assisted Large Neighborhood Search}

\author{Masashi Yamashita$^{1}$ and Shu Tanaka$^{1,2,3,4,5}$\thanks{shu.tanaka@keio.jp}}
\inst{$^1$Graduate School of Science and Technology, Keio University, Kanagawa
223-8522, Japan \\
$^2$Department of Applied Physics and Physico-Informatics, Keio University, Kanagawa 223-8522, Japan \\
$^3$Keio University Sustainable Quantum Artificial Intelligence Center (KSQAIC), Keio University, Tokyo 108-8345,
Japan \\
$^4$Human Biology-Microbiome-Quantum Research Center (WPI-Bio2Q), Keio University, Tokyo 108-8345, Japan \\
$^5$Green Computing System Research Organization, Waseda University, Shinjuku, Tokyo 162-0042, Japan
}

\abst{
Large-scale quadratic unconstrained binary optimization (QUBO) formulations of constrained combinatorial optimization problems often exceed the input-size limit of present Ising machines or suffer from degraded solution quality as the number of binary variables increases.
Large neighborhood search (LNS) mitigates this difficulty by sequentially optimizing restricted subproblems, but the structural factors that distinguish subproblems beyond the number of binary variables remain insufficiently characterized.
In this study, we examine vehicle routing problems and compare a construction based on the vehicle routes of the current solution, denoted by LNS-K, with a construction based on QUBO variables and constraint relations, denoted by LNS-Q, while controlling the number of binary variables in the subproblems.
Under the tested conditions, LNS-K obtained shorter total distances than LNS-Q in the matched-size comparisons, and the position variance, a measure of the spatial spread of the selected customers, decreased during the iterations in LNS-K.
These observations suggest that subproblem design for sequential optimization with Ising machines should consider not only subproblem size but also semantic and geometric structures inherited from the current solution.
}

\begin{document}
\maketitle

\section{Introduction}
\label{sec:introduction}

Large-scale constrained combinatorial optimization problems often require quadratic unconstrained binary optimization (QUBO) formulations with many binary variables.
Ising machines have been developed as hardware or software solvers for finding low-energy states of Ising models or QUBO models~\cite{lucas2014ising, kochenberger2014unconstrained, tanaka2017quantum, tanahashi2019application, zaman2021pyqubo, mohseni2022ising, chakrabarti2022quantum}.
They provide a computational approach to treating combinatorial optimization problems after an appropriate Ising or QUBO formulation is constructed.
However, current Ising machines have practical limitations when the number of binary variables is large.
Some machines have explicit input-size limitations.
Even when a large QUBO can be submitted, the solution quality can deteriorate as the number of variables and constraints increases.
This limitation is particularly relevant for constrained QUBO formulations, where the feasible region can occupy only a small fraction of the binary search space.

Large neighborhood search (LNS)~\cite{shaw1998using} related subproblem decomposition and reduction approaches provide a way to use an Ising machine for such large-scale problems~\cite{booth2017partitioning, qbsolv, karimi2017effective, karimi2017boosting, okada2019improving, nishimura2019item, okada2019efficient, irie2021hybrid, zhang2022solving, kikuchi2023hybrid, noguchi2023trip, hattori2025advantages, hattori2025impact, peng2025hybrid, ide2025extending}.
In LNS, a current solution is prepared first.
A subproblem is then constructed by selecting a subset of variables from the current solution, and the selected variables are reoptimized while the remaining variables are fixed.
By solving these smaller subproblems sequentially, the current solution can be improved without submitting the full QUBO to an Ising machine at every iteration.
The size of the subproblem is an important design parameter.
If the subproblem is too small, each update can modify only a limited part of the current solution.
If the subproblem is too large, the Ising machine may not solve the subproblem accurately.
Thus, there is a tradeoff between the range of possible updates and the solution accuracy of the subproblem.

Previous studies on Ising-machine-assisted LNS have often treated the number of binary variables in the subproblem as a primary control parameter.
This parameter is important because it directly determines the input size of the restricted QUBO.
However, the number of binary variables alone does not specify the structure of a subproblem.
Even with the same number of variables, different construction rules can select different sets of variables.
They can also produce different constraint relations, different route structures, and different geometric distributions of selected customers.
This observation leads to the central question of this study, namely whether the subproblem construction rule still affects the sequential optimization process when the number of binary variables in the subproblem is held fixed.

To address this question, we compare two subproblem construction rules for vehicle routing problems (VRPs)~\cite{toth2002vehicle}.
The first construction, denoted by LNS-K, is based on the route structure of the current solution.
It selects a subset of the vehicles and constructs a subproblem from those vehicles and the customers visited by them in the current solution.
The second construction, denoted by LNS-Q, is based on QUBO variables and their associated constraint relations.
It starts from the active binary variables of the current solution and constructs a subproblem by considering the related constraints.
This is not a comparison between a structured construction and one that ignores the problem structure.
LNS-Q also uses the QUBO representation and constraint relations.
The difference is that LNS-K directly uses vehicles and current routes as problem-domain units, whereas LNS-Q constructs subproblems from the QUBO representation.

The contributions of this paper are threefold.
First, we evaluate Ising-machine-assisted LNS and the naive approach of submitting the full QUBO once under matched total annealing-time budgets using a quantum annealing machine and a GPU-based Ising machine.
Second, we show that LNS-K and LNS-Q exhibit different final solution qualities and different iterative trajectories even when the numbers of binary variables in the subproblems are matched.
Third, we show that the customer sets selected by LNS-K become more spatially localized during the iterations, whereas those selected by LNS-Q remain nearly unchanged in their position variance.
These observations indicate that the number of binary variables alone is insufficient to characterize a subproblem.
They also suggest that structural information inherited from the current solution should be considered in subproblem design for sequential optimization with Ising machines.

The rest of this paper is organized as follows.
Section~\ref{sec:problem_formulation} formulates the VRP as a QUBO and defines the two subproblem construction rules, LNS-K and LNS-Q, together with the measures used in the comparison.
Section~\ref{sec:numerical_setup} describes the problem instances, the Ising machines, and the matched-size protocol.
Section~\ref{sec:results} presents the numerical results.
Section~\ref{sec:discussion} discusses their implications for subproblem design.
Section~\ref{sec:summary_conclusion} concludes the paper.
\section{Problem Formulation and Subproblem Construction}
\label{sec:problem_formulation}

In this section, we set up the problem and the subproblem construction rules used in the rest of the paper.
Section~\ref{subsec:restricted_qubo} describes how a restricted subproblem is obtained from a QUBO by fixing a subset of binary variables.
Section~\ref{subsec:vrp_formulation} formulates the vehicle routing problem as a QUBO, and Section~\ref{subsec:ising_lns} presents the large neighborhood search procedure.
Sections~\ref{subsec:lns_k} and \ref{subsec:lns_q} define the two subproblem construction rules, LNS-K and LNS-Q, and Section~\ref{subsec:measures} introduces the quantities used to compare them.

\subsection{Restricted QUBO for sequential optimization}
\label{subsec:restricted_qubo}

We first describe how a subproblem is obtained from a QUBO by fixing a subset of binary variables.
Let
\begin{equation}
    H(\boldsymbol{x}) =
    \boldsymbol{x}^{\mathsf T} Q \boldsymbol{x},
    \qquad
    \boldsymbol{x} \in \{0,1\}^{M},
    \label{eq:general_qubo}
\end{equation}
be a QUBO with a symmetric matrix \(Q\).
If the original QUBO matrix is not symmetric, we replace it by its symmetrized form without changing the value of the quadratic form.

At iteration \(m\), let \(S_m\) be the set of variables to be reoptimized, and let \(\bar{S}_m\) denote its complement.
The variables in \(\bar{S}_m\) are fixed to the current solution
\(\boldsymbol{x}^{(m)}_{\bar{S}_m}\).
The restricted QUBO submitted to an Ising machine is then written as
\begin{equation}
\begin{split}
    H_{S_m}
    \left(
        \boldsymbol{x}_{S_m}
        \mid
        \boldsymbol{x}^{(m)}_{\bar{S}_m}
    \right)
    &=
    \boldsymbol{x}_{S_m}^{\mathsf T}
    Q_{S_m,S_m}
    \boldsymbol{x}_{S_m}
    \\
    &\quad
    +2
    \boldsymbol{x}_{S_m}^{\mathsf T}
    Q_{S_m,\bar{S}_m}
    \boldsymbol{x}^{(m)}_{\bar{S}_m}
    + {\rm const}.
\end{split}
\label{eq:restricted_qubo}
\end{equation}
Here, the constant term does not affect the minimizer of the restricted QUBO.

Equation~\eqref{eq:restricted_qubo} shows that the number of free variables,
\(|S_m|\), is not the only property of a subproblem.
Even if two subproblems have the same number of binary variables, the internal interaction
\(Q_{S_m,S_m}\) and the effective linear term induced by the fixed variables can differ.
In this paper, Eq.~\eqref{eq:restricted_qubo} is used only to clarify this point, and we do not introduce additional structural measures such as boundary couplings or graph density.

\subsection{Vehicle routing problem}
\label{subsec:vrp_formulation}

We consider a vehicle routing problem (VRP) with one depot, \(N\) customers, and \(K\) vehicles.
The depot is indexed by \(0\), and the customers are indexed by
\(i=1,\ldots,N\).
The distance between sites \(i\) and \(j\) is denoted by \(d_{i,j}\).
Each vehicle starts from the depot, visits a subset of customers, and finally returns to the depot.
The objective is to minimize the total distance traveled by all vehicles.

We introduce a binary variable
\begin{equation}
    x^{(k)}_{t,i}
    =
    \begin{cases}
    1, & \text{if vehicle } k \text{ visits site } i
    \text{ at step } t,\\
    0, & \text{otherwise},
    \end{cases}
    \label{eq:vrp_binary_variable}
\end{equation}
where \(k=1,\ldots,K\), \(t=1,\ldots,T\), and \(i=0,\ldots,N\).
Each vehicle starts from the depot before the first step, and this initial depot visit is fixed rather than encoded by a binary variable.
The binary variables \(x^{(k)}_{t,i}\) therefore range over \(K\) vehicles, \(T\) steps, and \(N+1\) sites, so the number of binary variables in the original problem is
\begin{equation}
    K(N+1)T .
    \label{eq:number_original_variables}
\end{equation}

The total distance is written as
\begin{equation}
\begin{split}
    D(\boldsymbol{x})
    &=
    \sum_{k=1}^{K}
    \left[
    \sum_{i=0}^{N}
    d_{0,i} x^{(k)}_{1,i}
    +
    \sum_{t=1}^{T-1}
    \sum_{i=0}^{N}
    \sum_{j=0}^{N}
    d_{i,j}
    x^{(k)}_{t,i}
    x^{(k)}_{t+1,j}
    \right.
    \\
    &\qquad\qquad
    \left.
    +
    \sum_{i=0}^{N}
    d_{i,0} x^{(k)}_{T,i}
    \right].
\end{split}
\label{eq:total_distance}
\end{equation}
The first and last terms represent the departure from and the return to the depot, respectively.

The feasible routes must satisfy the following constraints.
First, every customer must be visited exactly once,
\begin{equation}
    \sum_{k=1}^{K}
    \sum_{t=1}^{T}
    x^{(k)}_{t,i}
    =
    1,
    \qquad
    i=1,\ldots,N .
    \label{eq:constraint_customer_once}
\end{equation}
Second, each vehicle must visit exactly one site at each step,
\begin{equation}
    \sum_{i=0}^{N}
    x^{(k)}_{t,i}
    =
    1,
    \qquad
    k=1,\ldots,K,\quad
    t=1,\ldots,T .
    \label{eq:constraint_one_site}
\end{equation}
Third, once a vehicle returns to the depot before the final step, it stays at the depot,
\begin{equation}
    x^{(k)}_{t,0}
    \left(1-x^{(k)}_{t+1,0}\right)
    =
    0,
    \qquad
    k=1,\ldots,K,\quad
    t=1,\ldots,T-1 .
    \label{eq:constraint_depot_stay}
\end{equation}
When $KT$ is equal to $N$, any solution that satisfies Eqs.~\eqref{eq:constraint_customer_once} and \eqref{eq:constraint_one_site} contains no depot visit at any step, so Eq.~\eqref{eq:constraint_depot_stay} is automatically satisfied.
Using penalty coefficients
\(\mu_{\mathrm A}\), \(\mu_{\mathrm B}\), and \(\mu_{\mathrm C}\),
the QUBO for the VRP is written as
\begin{equation}
\begin{split}
    H_{\rm VRP}(\boldsymbol{x})
    &=
    D(\boldsymbol{x})
    +\mu_{\mathrm A}
    \sum_{i=1}^{N}
    \left(
    1-
    \sum_{k=1}^{K}
    \sum_{t=1}^{T}
    x^{(k)}_{t,i}
    \right)^2
    \\
    &\quad
    +\mu_{\mathrm B}
    \sum_{k=1}^{K}
    \sum_{t=1}^{T}
    \left(
    1-
    \sum_{i=0}^{N}
    x^{(k)}_{t,i}
    \right)^2
    \\
    &\quad
    +\mu_{\mathrm C}
    \sum_{k=1}^{K}
    \sum_{t=1}^{T-1}
    x^{(k)}_{t,0}
    \left(1-x^{(k)}_{t+1,0}\right).
\end{split}
\label{eq:vrp_qubo}
\end{equation}
This QUBO can be written in the general form of Eq.~\eqref{eq:general_qubo}.

\subsection{Ising-machine-assisted large neighborhood search}
\label{subsec:ising_lns}

We use large neighborhood search (LNS) to improve a current feasible solution.
At each iteration, a subset of variables is selected to form a subproblem, and these variables are reoptimized by an Ising machine while the remaining variables are fixed to the current solution.
The solution returned from the subproblem is then embedded back into the original variable space to construct a candidate solution.
Algorithm~\ref{alg:lns} summarizes the procedure.
The construction of the feasible initial current solution used in the experiments is described in Sec.~\ref{subsec:problem_instances}.
The acceptance rule is conservative, so the current solution is updated only when the reconstructed candidate solution is feasible and has a shorter total distance than the current solution.

\begin{algorithm}[tbp]
\caption{Ising-machine-assisted large neighborhood search}
\label{alg:lns}
\begin{algorithmic}[1]
\State Generate a feasible initial current solution \(\boldsymbol{x}^{(0)}\).
\For{\(m=0,1,\ldots,M_{\rm iter}-1\)}
    \State Construct a variable set \(S_m\) from the current solution \(\boldsymbol{x}^{(m)}\).
    \State Build the restricted QUBO \(H_{S_m}\!\left(\boldsymbol{x}_{S_m} \mid \boldsymbol{x}^{(m)}_{\bar{S}_m}\right)\).
    \State Solve the restricted QUBO with an Ising machine.
    \State Reconstruct a candidate solution \(\tilde{\boldsymbol{x}}\) for the original problem.
    \If{\(\tilde{\boldsymbol{x}}\) is feasible and \(D(\tilde{\boldsymbol{x}})<D(\boldsymbol{x}^{(m)})\)}
        \State \(\boldsymbol{x}^{(m+1)} \gets \tilde{\boldsymbol{x}}\)
    \Else
        \State \(\boldsymbol{x}^{(m+1)} \gets \boldsymbol{x}^{(m)}\)
    \EndIf
\EndFor
\State \textbf{return} the final current solution \(\boldsymbol{x}^{(M_{\rm iter})}\).
\end{algorithmic}
\end{algorithm}

The feasibility of the accepted current solution should be distinguished from the feasibility of the raw samples returned by the Ising machine.
In the LNS procedure, a raw sample is first used to reconstruct a candidate solution for the original problem, and the candidate is accepted only when it satisfies the VRP constraints and improves the total distance.
The concrete reconstruction rule used in the experiments is described in Sec.~\ref{subsec:ising_machines}.
Otherwise the candidate is rejected and the current solution is retained.
Because the initial current solution is feasible and every accepted candidate is feasible, the sequence of accepted current solutions remains in the feasible region of the original problem.
The feasibility of this accepted sequence therefore reflects the reconstruction and acceptance rule in addition to the sampling behavior of the Ising machine.

\subsection{Route-structure-based construction, LNS-K}
\label{subsec:lns_k}

We call the route-structure-based construction LNS-K, which uses the route structure of the current solution in the VRP.
At iteration \(m\), LNS-K selects a subset
\(\mathcal{K}_m\) of \(K'\) vehicles uniformly at random from the \(K\) vehicles.
The customer set included in the subproblem is then defined by
\begin{align}
    \mathcal{C}^{\rm K}_m
    =
    \left\{
    i \in \{1,\ldots,N\}
    \mid
    x^{(k),(m)}_{t,i}=1
    \text{ for some }
    k \in \mathcal{K}_m
    \right. \nonumber \\
    \left.
    \text{ and }
    t \in \{1,\ldots,T\}
    \right\}.
    \label{eq:customer_set_lnsk}
\end{align}
The subproblem consists of the selected vehicles and the customers in
\(\mathcal{C}^{\rm K}_m\), together with the depot.

The free variables in LNS-K are
\begin{equation}
    S^{\rm K}_m
    =
    \left\{
    (k,t,i)
    \mid
    k \in \mathcal{K}_m,\quad
    t=1,\ldots,T,\quad
    i \in \mathcal{C}^{\rm K}_m \cup \{0\}
    \right\},
    \label{eq:variables_lnsk}
\end{equation}
and the variables in the complement of \(S^{\rm K}_m\) are fixed to the current solution.
Because the same VRP constraints are imposed only on the selected vehicles and the selected customers, any feasible solution of the LNS-K subproblem can be embedded into the original VRP without changing the routes of the unselected vehicles and without violating the original constraints.
Consequently, LNS-K moves the current solution within the feasible region at the level of accepted reconstructed solutions.

A key feature of LNS-K is that the subproblem is constructed from vehicles and current routes, so the unit of variable selection is a semantic unit of the original VRP.
This subsection defines the construction rule only, and its performance is evaluated in Sec.~\ref{sec:results}.

\subsection{QUBO-variable-based construction, LNS-Q}
\label{subsec:lns_q}

We call the QUBO-variable-based construction LNS-Q, which uses the QUBO variables and the constraint relations in the QUBO representation.
At iteration \(m\), we first consider the set of active variables in the current solution,
\begin{equation}
    \mathcal{A}_m
    =
    \left\{
    (k,t,i)
    \mid
    x^{(k),(m)}_{t,i}=1,\quad i\neq 0
    \right\}.
    \label{eq:active_variables}
\end{equation}
LNS-Q selects \(n'\) active variables from \(\mathcal{A}_m\) uniformly at random, and these selected variables are used as anchors.
The binary variables that appear in the constraints associated with the anchors are fixed to the current solution, and the remaining binary variables form the subproblem variable set \(S^{\rm Q}_m\).
Because a larger \(n'\) fixes more variables through the associated constraints, the number of free binary variables in the subproblem decreases as \(n'\) increases.
This dependence lets \(n'\) vary the subproblem size more finely than \(K'\) does in LNS-K.

LNS-Q does not ignore the constraint structure, since it uses the QUBO variables and the constraint relations induced by the QUBO formulation.
The difference from LNS-K is that LNS-Q does not directly use vehicles, routes, or spatial customer locations as units for subproblem generation.

\subsection{Measures used in the comparison}
\label{subsec:measures}

We compare LNS-K and LNS-Q using quantities that are directly obtained from the numerical data.
The first quantity is the number of binary variables in the subproblem,
\begin{equation}
    B_m = |S_m|.
    \label{eq:number_subproblem_variables}
\end{equation}
For the matched-size comparison in Sec.~\ref{sec:results}, the parameters \(K'\) and \(n'\) are chosen so that LNS-K and LNS-Q have the same value of \(B_m\).

The second quantity is the total distance of the current solution,
\begin{equation}
    D^{(m)} = D(\boldsymbol{x}^{(m)}),
    \label{eq:current_distance}
\end{equation}
where \(D(\boldsymbol{x})\) is defined in Eq.~\eqref{eq:total_distance}.
We use \(D^{(m)}\) to analyze the iterative improvement process and \(D^{(M_{\rm iter})}\) to compare the final solution quality.

The third quantity is the position variance of the selected customer set.
Let \(\mathcal{C}_m\) be the customer set included in the subproblem at iteration \(m\), and for a customer \(i\), let \((u_i,v_i)\) be its two-dimensional coordinates.
We define the mean position
\begin{equation}
    \bar{u}_m
    =
    \frac{1}{|\mathcal{C}_m|}
    \sum_{i\in\mathcal{C}_m}
    u_i,
    \qquad
    \bar{v}_m
    =
    \frac{1}{|\mathcal{C}_m|}
    \sum_{i\in\mathcal{C}_m}
    v_i ,
    \label{eq:mean_position}
\end{equation}
and the position variance
\begin{equation}
    V_{\rm pos}^{(m)}
    =
    \frac{1}{|\mathcal{C}_m|}
    \sum_{i\in\mathcal{C}_m}
    \left(u_i-\bar{u}_m\right)^2
    +
    \frac{1}{|\mathcal{C}_m|}
    \sum_{i\in\mathcal{C}_m}
    \left(v_i-\bar{v}_m\right)^2.
    \label{eq:position_variance}
\end{equation}
The depot is not included in \(\mathcal{C}_m\) when computing this quantity.

The position variance is used as a descriptor of the geometric spread of the selected customer set, and a smaller value means that the selected customers are more spatially localized.
We do not define the position variance as a causal measure of the improvement in the total distance.

\section{Numerical Setup}
\label{sec:numerical_setup}

This section describes the numerical experiments.
Section~\ref{subsec:problem_instances} specifies the VRP instances.
Section~\ref{subsec:ising_machines} describes the two Ising machines and the annealing conditions.
Section~\ref{subsec:roles_of_machines} explains the different roles of the two machines, and Section~\ref{subsec:matched_size_protocol} defines the matched-size protocol used to compare LNS-K and LNS-Q.

\subsection{Problem instances}
\label{subsec:problem_instances}

We used vehicle routing problem (VRP) instances with one depot and $N$ customers.
The depot was placed at the center of the unit square.
The coordinates of the customers were sampled independently from the uniform distribution on the unit square.
For each value of $N$, the customer coordinates were sampled once and then fixed, so a single problem instance was used for each value of $N$.
The distance between two sites was defined by the Euclidean distance.
We denote by $K$ the number of vehicles and by $T$ the number of steps assigned to each vehicle, so that the original problem has $K(N+1)T$ binary variables, as defined in Sec.~\ref{subsec:vrp_formulation}.

The feasible initial current solution in Algorithm~\ref{alg:lns} was generated randomly as follows.
The $N$ customers were permuted uniformly at random.
The permuted customers were then assigned to the $K$ vehicles cyclically, so that the numbers of customers assigned to the vehicles differ by at most one.
Each vehicle visits its assigned customers in the assigned order from the first step.
When the number of assigned customers is smaller than $T$, the vehicle returns to the depot after visiting all of its assigned customers and stays at the depot in the remaining steps.
In all instances in Table \ref{tab:numerical_setup}, the number of customers assigned to each vehicle does not exceed $T$, so this construction always yields an initial solution that satisfies Eqs.~\eqref{eq:constraint_customer_once}, \eqref{eq:constraint_one_site}, and \eqref{eq:constraint_depot_stay}.
When $N$ is equal to $KT$, as in all the Fixstars Amplify AE instances, every vehicle visits exactly $T$ customers.
This construction uses neither the distances nor the customer coordinates, so the initial assignment of customers to vehicles is statistically independent of the customer locations.

Table~\ref{tab:numerical_setup} summarizes the problem instances and annealing conditions.
For each problem setting, 30 initial solutions were generated by repeating this procedure with independent random permutations.
The small instances with $N=5$, 6, and 7 were used for the D-Wave Advantage 4.1 experiments.
The larger instances with $N=100$, 200, 300, and 400 were used for the Fixstars Amplify AE experiments.
The 400-customer instance was used for the detailed comparison between LNS-K and LNS-Q.

\begin{table*}[tbp]
\centering
\caption{Problem instances and annealing conditions used in this study. For each value of $N$, the customer coordinates were sampled uniformly at random on the unit square with the depot at the center and then fixed, and 30 initial solutions were used. In (a), the rightmost column lists the common value of the penalty coefficients $\mu_\mathrm{A}$ and $\mu_\mathrm{B}$ for each instance, and $\mu_\mathrm{C}=0$ in all cases.}
\label{tab:numerical_setup}
\small

(a) Problem instances.

\begin{tabular}{lcccc}
\hline
Ising machine & $N$ & $K$ & $T$ & $\mu_\mathrm{A}=\mu_\mathrm{B}$\\
\hline
D-Wave Advantage 4.1 & 5   & 3  & 3 & 0.73\\
D-Wave Advantage 4.1 & 6   & 3  & 3 & 0.79\\
D-Wave Advantage 4.1 & 7   & 3  & 4 & 0.53\\
Fixstars Amplify AE  & 100 & 10 & 10 & 1.21\\
Fixstars Amplify AE  & 200 & 10 & 20 & 1.30\\
Fixstars Amplify AE  & 300 & 10 & 30 & 1.32\\
Fixstars Amplify AE  & 400 & 10 & 40 & 1.32\\
\hline
\end{tabular}

\vspace{2mm}

(b) Annealing conditions for the comparison between LNS and the naive approach. The total annealing time is the product of the number of calls, the number of reads per call, and the annealing time per read.

\begin{tabular}{llcccc}
\hline
Ising machine & Algorithm & Number of calls & Number of reads per call & Annealing time per call & Total annealing time \\
\hline
D-Wave Advantage 4.1 & LNS   & 100 & 400 & $20~\mu{\rm s}$    & $800~{\rm ms}$ \\
D-Wave Advantage 4.1 & Naive & 1   & 400 & $2000~\mu{\rm s}$  & $800~{\rm ms}$ \\
Fixstars Amplify AE  & LNS   & 60  & 1 & $10~{\rm s}$       & $600~{\rm s}$ \\
Fixstars Amplify AE  & Naive & 1   & 1 & $600~{\rm s}$      & $600~{\rm s}$ \\
\hline
\end{tabular}
\end{table*}

\subsection{Ising machines and annealing conditions}
\label{subsec:ising_machines}

We used two Ising machines.
D-Wave Advantage 4.1 was used as a quantum annealing machine, and Fixstars Amplify AE was used as a GPU-based Ising machine.

Both Ising machines were accessed through the Fixstars Amplify SDK, and version 0.11.1 of the SDK was used.
For D-Wave Advantage 4.1, the solver was specified as Advantage\_system4.1, and the annealing time and the number of reads were specified explicitly as described below.
Minor embedding onto the hardware graph was performed by the automatic embedding of the SDK.

The chain strength, the chain break resolution method, the automatic rescaling of the coefficients, the spin reversal transforms, and the annealing schedule were not specified and were left at the default settings of the SDK.
For Fixstars Amplify AE, the annealing time was specified as described below, and the other settings were left at the default settings of the SDK.
The penalty coefficients $\mu_\mathrm{A}$ and $\mu_\mathrm{B}$ in Eq.~\eqref{eq:vrp_qubo} were set to the maximum element of the distance matrix of each instance, namely $\mu_\mathrm{A} = \mu_\mathrm{B} = \max_{i,j} d_{i,j}$.
The same rule was applied to all instances without further tuning, and the same values were used in the naive approach and in all restricted QUBOs in LNS.
The resulting values are listed in Table~\ref{tab:numerical_setup}.
The penalty coefficient $\mu_\mathrm{C}$ was set to zero, so the third penalty term in Eq.~\eqref{eq:vrp_qubo} was not included in the implemented QUBO.
For the Fixstars Amplify AE instances, $T$ is equal to $N/K$, so $KT$ is equal to $N$ and Eq.~\eqref{eq:constraint_depot_stay} is automatically satisfied by any solution that satisfies Eqs.~\eqref{eq:constraint_customer_once} and \eqref{eq:constraint_one_site}, as noted in Sec.~\ref{subsec:vrp_formulation}.
For the D-Wave Advantage 4.1 instances, $KT$ is larger than $N$, and a solution satisfying Eqs.~\eqref{eq:constraint_customer_once} and \eqref{eq:constraint_one_site} can contain depot visits before the final step.
Such depot visits are removed in the reconstruction step.
For each vehicle, the customers in the step sequence are packed forward in their original order, and the depot occupies the remaining trailing steps.
The reconstructed solution therefore satisfies Eq.~\eqref{eq:constraint_depot_stay}, and Eq.~\eqref{eq:constraint_depot_stay} is enforced by the reconstruction step rather than by the penalty term for these instances.
By the triangle inequality, this reconstruction does not increase the total distance.
The total distances reported in this paper are those of the reconstructed solutions, and the same reconstruction was applied to the solutions obtained by the naive approach.
This choice of the penalty coefficients does not guarantee that the lowest energy state of the submitted QUBO is feasible, and the feasibility of the reported solutions is ensured by the feasibility check in the sample selection and by the acceptance rule in Sec.~\ref{subsec:ising_lns}.

For the comparison between LNS and the naive approach, the total annealing time was matched.
In the naive approach, the full QUBO problem was submitted to an Ising machine once.
In LNS, smaller subproblems were submitted sequentially, and the current solution was updated after each subproblem solution.

For D-Wave Advantage 4.1, the number of reads per call was set to 400 for both LNS and the naive approach.
LNS was run for 100 iterations with an annealing time of $20~\mu{\rm s}$ per read, whereas the naive approach used one call with an annealing time of $2000~\mu{\rm s}$ per read.
The total annealing time, namely the product of the annealing time per read, the number of reads per call, and the number of calls, was therefore 800~ms for both approaches.

For Fixstars Amplify AE, each call returned a single solution.
LNS was run for 60 iterations with an annealing time of 10~s per call, whereas the naive approach used one call with an annealing time of 600~s.
The total annealing time was 600~s for both approaches.

In each call to D-Wave Advantage 4.1, samples violating the constraints were not filtered out, and the returned samples were sorted in ascending order of energy.
The sample used to reconstruct the candidate solution was the lowest energy sample among those satisfying the constraints corresponding to Eqs.~\eqref{eq:constraint_customer_once} and \eqref{eq:constraint_one_site}.
In LNS, when none of the 400 samples satisfied the constraints, the candidate was discarded and the current solution was retained in that iteration.
In the naive approach, the reported solution was likewise the lowest energy feasible sample among the 400 samples of the single call.

The matched quantity in this comparison is the total annealing time.
It does not include the time required for QUBO construction, communication with the machine, minor embedding, decoding, or postprocessing.
Therefore, the results should be interpreted as a comparison under a controlled annealing-time budget.
It should not be interpreted as an end-to-end wall-clock-time comparison.
For reference, in the D-Wave Advantage 4.1 experiments for the instance with $N = 5$ and $T = 3$, the total QPU access time was 9.87~s for LNS and 0.86~s for the naive approach, whereas the total annealing time was 800~ms for both.
This difference mainly reflects the programming and readout overheads that accumulate over the calls and reads.

\subsection{Roles of the two Ising machines}
\label{subsec:roles_of_machines}

The two Ising machines play different roles in this study.
The D-Wave Advantage 4.1 results are used to examine the behavior of sequential subproblem optimization for small constrained VRP instances.
They provide a small-scale test of the LNS framework with a quantum annealing machine.
In the D-Wave Advantage 4.1 experiments, the subproblems in LNS were constructed by LNS-K with $K' = 2$, and LNS-Q was not applied to the small instances.
Because the number of customers served by the two selected vehicles changes during the iterations, the number of binary variables in the subproblem also changes.
For the instances with $(N, T) = (5, 3)$, $(6, 3)$, and $(7, 4)$, the number of binary variables in the subproblem ranged from 18 to 36, from 24 to 42, and from 32 to 64, with averages of 26.2, 30.3, and 45.1, respectively, whereas the numbers of binary variables in the original problems are 54, 63, and 96.
In contrast, for the Fixstars Amplify AE instances, $T$ is equal to $N/K$ and every vehicle serves exactly $T$ customers in any feasible solution, so the number of binary variables in the LNS-K subproblem is constant during the iterations.

The Fixstars Amplify AE results are used for larger VRP instances.
They are used first to compare LNS with the naive approach for $N=100$, 200, 300, and 400.
They are then used for the detailed comparison between LNS-K and LNS-Q.

We do not use these results to compare the hardware performance of D-Wave Advantage 4.1 and Fixstars Amplify AE.
The two machines were applied to different problem sizes and used for different purposes.
We also do not claim that the effect of subproblem construction is confirmed on both hardware platforms.
The structure comparison between LNS-K and LNS-Q in Secs.~\ref{subsec:dependence_subproblem_construction}--\ref{subsec:geometric_evolution} is based on the large-scale Fixstars Amplify AE experiments.

\subsection{Matched-size protocol}
\label{subsec:matched_size_protocol}

The detailed comparison between LNS-K and LNS-Q was performed for the 400-customer instance.
For this instance, the original QUBO contains
\begin{equation}
    K(N+1)T = 10 \times 401 \times 40 = 160400
\end{equation}
binary variables.

In LNS-K, the subproblem size is controlled by the number $K'$ of selected vehicles, and in LNS-Q it is controlled by the parameter $n'$ used in the QUBO-variable-based construction.
For the matched-size comparison in Fig.~\ref{fig:totaldistances_positionvariance_lns_k_and_q}, we chose pairs of $K'$ and $n'$ so that LNS-K and LNS-Q have the same number of binary variables in the subproblem.
Table~\ref{tab:matched_subproblem_sizes} lists the matched pairs, and this protocol lets us compare the iterative behavior of the two construction rules under a fixed number of binary variables.
The same set of 30 initial solutions was shared between LNS-K and LNS-Q.
Therefore, in every matched-size comparison in Sec.~\ref{sec:results}, the two methods start from identical initial current solutions.

\begin{table}[tbp]
\centering
\caption{Matched subproblem sizes for LNS-K and LNS-Q.}
\label{tab:matched_subproblem_sizes}
\small
\begin{tabular}{ccc}
\hline
$K'$ in LNS-K & $n'$ in LNS-Q & Matched number of binary variables \\
\hline
2 & 320 & 6480 \\
3 & 280 & 14520 \\
4 & 240 & 25760 \\
5 & 200 & 40200 \\
\hline
\end{tabular}
\end{table}

In Sec.~\ref{sec:results}, we analyze the final total distance, the total distance of the current solution during the iterations, and the position variance of the selected customer set.
\section{Results}
\label{sec:results}

This section presents the numerical results.
Section~\ref{subsec:equal_annealing_budget} compares LNS with the naive approach under an equal annealing-time budget.
Section~\ref{subsec:dependence_subproblem_construction} examines how the final solution quality depends on the subproblem construction rule.
Section~\ref{subsec:iterative_improvement} compares the iterative improvement processes at matched subproblem sizes, and Section~\ref{subsec:geometric_evolution} analyzes the geometric evolution of the selected customer sets.

\subsection{Sequential optimization under an equal annealing budget}
\label{subsec:equal_annealing_budget}

\begin{figure}[hbtp]
    \centering
    \begin{minipage}[t]{0.49\linewidth}
        \centering
        \includegraphics[width=\linewidth]{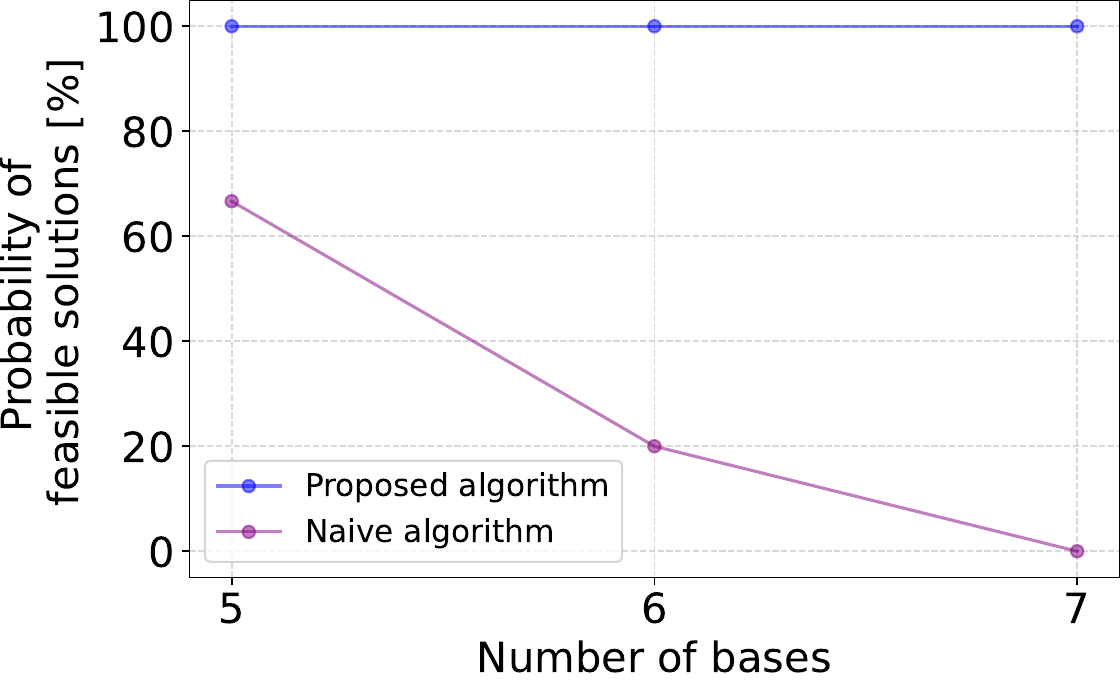}\\
        (a)
    \end{minipage}
    \begin{minipage}[t]{0.49\linewidth}
        \centering
        \includegraphics[width=\linewidth]{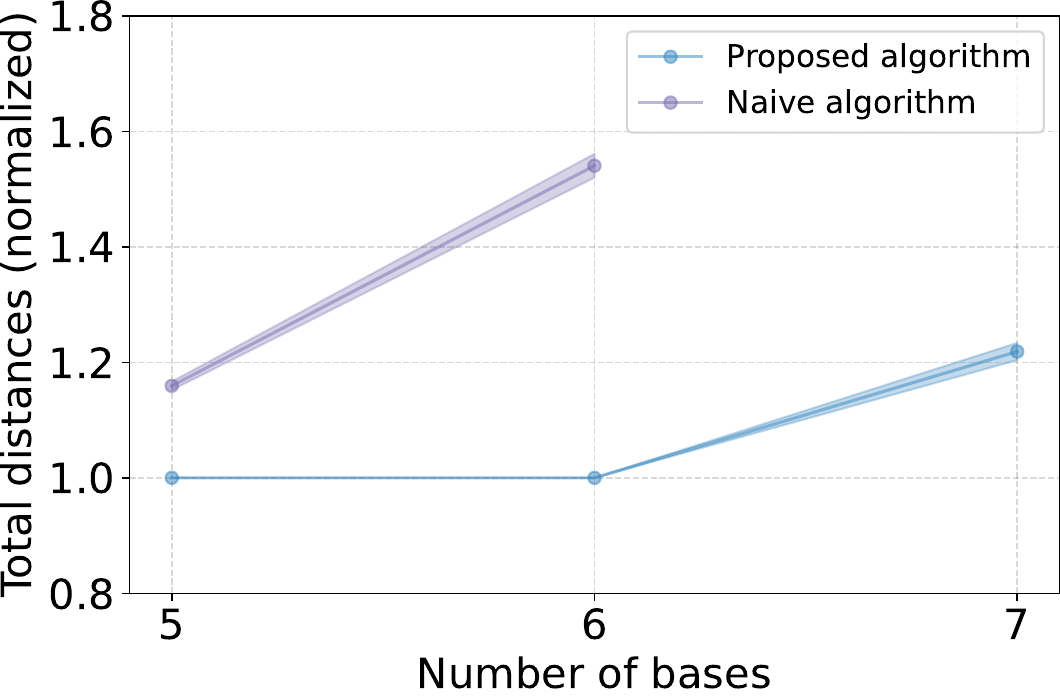}\\
        (b)
    \end{minipage}\\
    \vspace{2mm}
    \begin{minipage}[t]{0.49\linewidth}
        \centering
        \includegraphics[width=\linewidth]{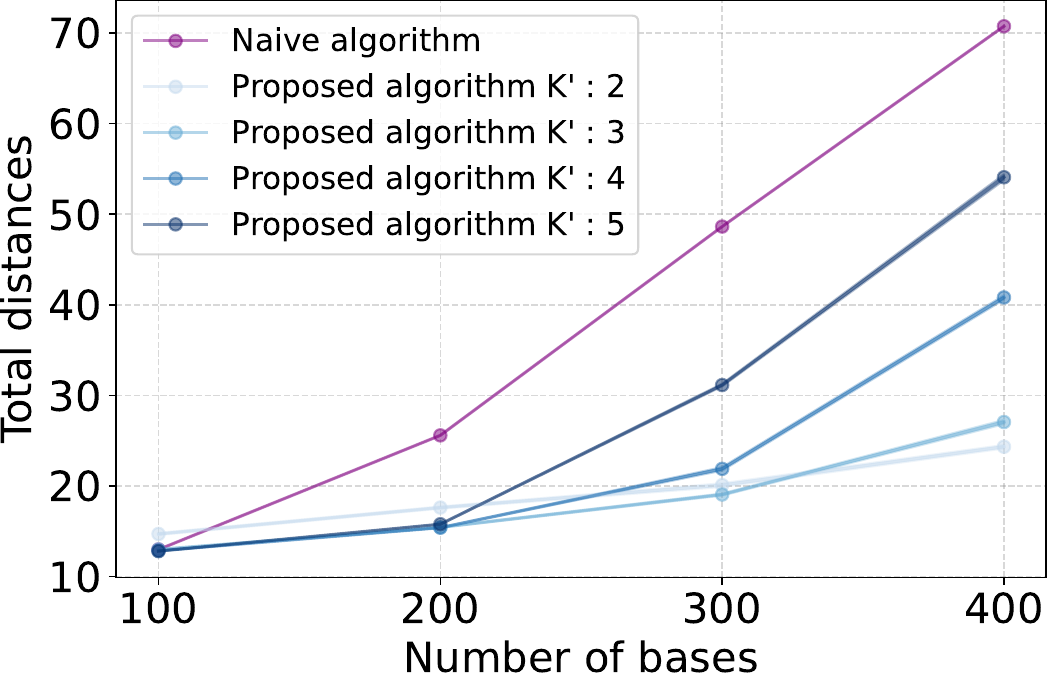}\\
        (c)
    \end{minipage}
    \begin{minipage}[t]{0.49\linewidth}
        \centering
        \includegraphics[width=\linewidth]{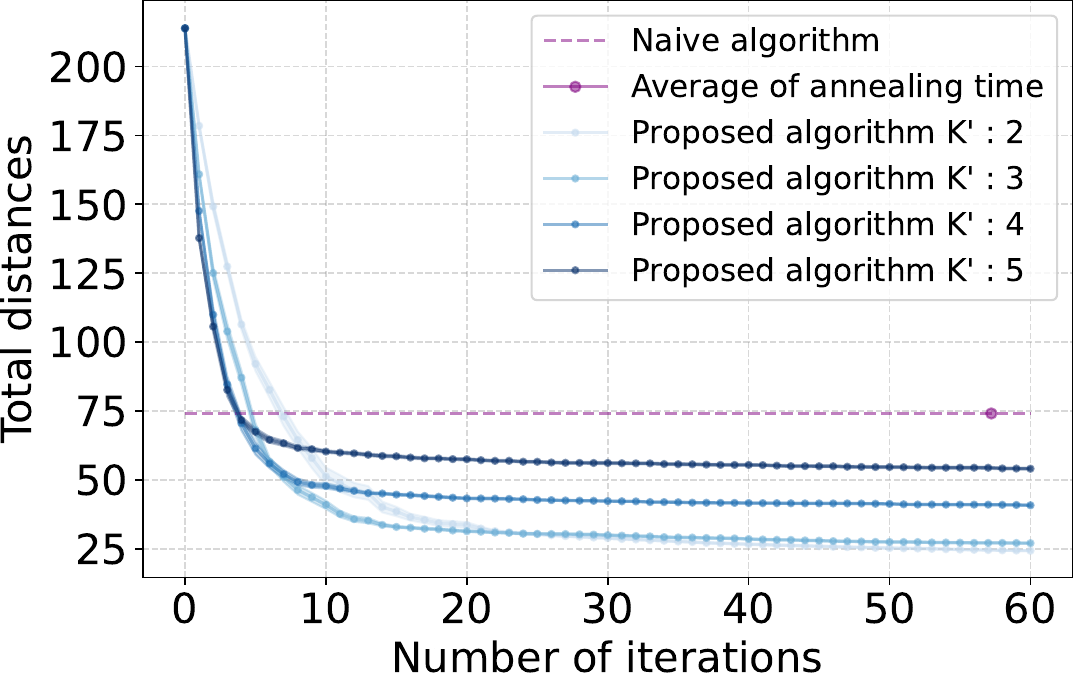}\\
        (d)
    \end{minipage}
    \caption{
Comparison between LNS and the naive approach under an equal total annealing-time budget.
(a) Feasible-solution probability obtained with D-Wave Advantage 4.1 for small VRP instances.
(b) Normalized total distance obtained with D-Wave Advantage 4.1.
(c) Final total distance obtained with Fixstars Amplify AE for VRP instances with 100, 200, 300, and 400 customers.
(d) Total distance of the current solution as a function of the iteration number for the 400-customer instance.
In panels (a), (b), and (c), the horizontal axis shows the number of customers $N$.
In the naive approach, the full QUBO problem is submitted to the Ising machine once.
In LNS, subproblems are solved sequentially and the current solution is updated when the reconstructed solution is feasible and improves the total distance.
The total annealing time is matched between LNS and the naive approach.
The label Proposed algorithm in the legends denotes LNS-K.
In (a) and (b), the instances are $(N, T) = (5, 3), (6, 3),$ and $(7, 4)$ with $K = 3$, and the subproblems are constructed with $K' = 2$.
In (c) and (d), the values of $K'$ are indicated in the legends.
}    
\label{fig:equal_annealing_budget}
\end{figure}

Before comparing different subproblem construction rules, we first examine the effect of sequential subproblem optimization under an equal annealing-time budget.
Figure~\ref{fig:equal_annealing_budget} compares LNS with the naive approach, in which the full QUBO problem is directly submitted to an Ising machine.
The total annealing time is matched between the two approaches.
For D-Wave Advantage 4.1, LNS uses 100 iterations with an annealing time of $20~\mu{\rm s}$ per read, whereas the naive approach uses one call with an annealing time of $2000~\mu{\rm s}$ per read, and both use 400 reads per call.
For Fixstars Amplify AE, LNS uses 60 iterations with an annealing time of $10~{\rm s}$ per call, whereas the naive approach uses one call with an annealing time of $600~{\rm s}$.

Figures~\ref{fig:equal_annealing_budget}(a) and \ref{fig:equal_annealing_budget}(b) show the results obtained with D-Wave Advantage 4.1 for small VRP instances.
In Fig.~\ref{fig:equal_annealing_budget}(a), the feasible-solution probability of the naive approach is defined as the fraction of the 30 runs in which at least one of the 400 samples satisfies Eqs.~\eqref{eq:constraint_customer_once} and ~\eqref{eq:constraint_one_site}.
For each such run, the reported solution is the lowest energy sample among the samples satisfying Eqs.~\eqref{eq:constraint_customer_once} and ~\eqref{eq:constraint_one_site}, followed by the reconstruction described in Sec.~\ref{subsec:ising_machines}.
For LNS, the feasible-solution probability is 100\% by construction, because the initial solution is feasible and the current solution is updated only when the reconstructed candidate is feasible.
As noted in Sec.~\ref{subsec:ising_lns}, this does not mean that the raw samples returned by the Ising machine are always feasible.
The feasible-solution probability of the naive approach decreases as the problem size increases.
This comparison should be interpreted with care because LNS includes a reconstruction and acceptance step for the current solution.
We therefore use this result only as evidence that the sequential optimization procedure can operate on constrained small instances.

In Fig.~\ref{fig:equal_annealing_budget}(b), the total distance is normalized by the exact optimal total distance of each instance.
The exact optimal values were obtained by solving a mixed integer programming formulation of the same VRP with the CBC solver and were verified by exhaustive enumeration of all feasible route sets.
The exact optimal total distances are 1.813, 2.327, and 1.920 for the instances with $N = 5, 6$, and $7$, respectively.
The same values are used for LNS and the naive approach.
For the naive approach, the result is shown only when at least one feasible solution was obtained, so no naive result is shown for the instance with $N = 7$.
Figure~\ref{fig:equal_annealing_budget}(b) also shows that, among the feasible solutions obtained in the tested instances, LNS gives shorter normalized total distances than the naive approach.
Thus, under the same total annealing-time budget, solving smaller subproblems sequentially improves the solution quality for the small instances tested on D-Wave Advantage 4.1.

Figures~\ref{fig:equal_annealing_budget}(c) and \ref{fig:equal_annealing_budget}(d) show the corresponding results obtained with Fixstars Amplify AE.
Figure~\ref{fig:equal_annealing_budget}(c) plots the final total distance for VRP instances with 100, 200, 300, and 400 customers.
The difference between LNS and the naive approach becomes more visible as the problem size increases.
For the 400-customer instance, LNS gives a much shorter final total distance than the naive approach.

Figure~\ref{fig:equal_annealing_budget}(d) shows the iterative behavior for the 400-customer instance.
The total distance of the current solution decreases rapidly during the early iterations.
This distance drops below the final total distance obtained by the naive approach well before the full 60-iteration budget is consumed.

These results support the use of Ising-machine-assisted LNS as a sequential optimization framework.
The following sections focus on the main question of this study, namely how the construction rule of the subproblem affects the optimization behavior when the subproblem size is controlled.

\subsection{Dependence on subproblem construction}
\label{subsec:dependence_subproblem_construction}

\begin{figure}[hbtp]
    \centering
    \includegraphics[width=\linewidth]{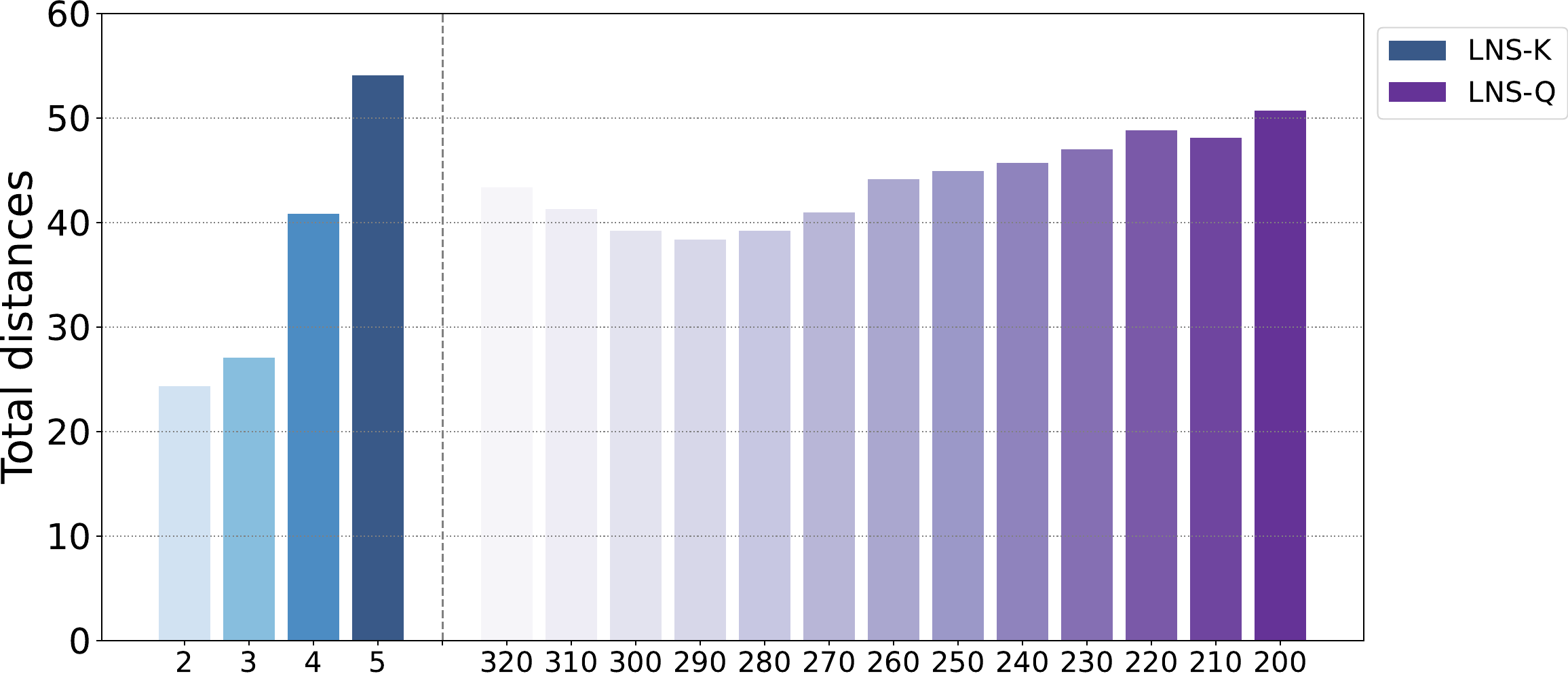}
    \caption{Final total distance obtained by LNS-K and LNS-Q for a vehicle routing problem with 400 customers. LNS-K constructs a subproblem from selected vehicles and the customers visited by those vehicles in the current solution, whereas LNS-Q constructs a subproblem from QUBO variables and the associated constraints. The horizontal labels indicate $K'$ for LNS-K and $n'$ for LNS-Q. The corresponding numbers of binary variables in the subproblems are summarized in Table~\ref{tab:matched_subproblem_sizes}. Although $n'$ allows LNS-Q to control the subproblem size more finely, LNS-K gives shorter final total distances under the tested conditions.}
    \label{fig:totaldistances_lns_k_and_q}
\end{figure}

We first examine the dependence of the final solution quality on the subproblem construction rule.
Figure~\ref{fig:totaldistances_lns_k_and_q} shows the final total distance obtained by LNS-K and LNS-Q for the vehicle routing problem with 400 customers.
In LNS-K, the subproblem size is controlled by the number $K'$ of selected vehicles.
In LNS-Q, the subproblem size is controlled by the parameter $n'$.
The corresponding numbers of binary variables in the subproblems are summarized in Table~\ref{tab:matched_subproblem_sizes}.

The LNS-Q results provide a finer scan of the subproblem size than the LNS-K results.
The final total distance changes with $n'$ in LNS-Q.
However, the best final distances obtained by LNS-K are shorter than those obtained by LNS-Q in the tested range of $n'$.
In particular, the LNS-Q data include subproblem sizes that are the same as those of LNS-K for several values of $K'$.
For these matched or comparable subproblem sizes, the final total distance still depends on the construction rule of the subproblem.

These results show that the number of binary variables in the subproblem is not sufficient to specify the behavior of the sequential optimization.
The construction rule used to choose the variables in the subproblem also affects the final total distance.

Among the tested values of $K'$ in LNS-K, $K'=2$ gives the shortest final total distance.
This value corresponds to the smallest vehicle-based subproblem size examined in LNS-K.
The result indicates that, under the present experimental conditions, the best LNS-K result in the tested parameter range is obtained at the lower end of the vehicle-based subproblem-size range.

\subsection{Iterative improvement at matched subproblem sizes}
\label{subsec:iterative_improvement}

\begin{figure}[hbtp]
    \centering
    \begin{minipage}[t]{0.49\linewidth}
        \centering
        \includegraphics[width=\linewidth]{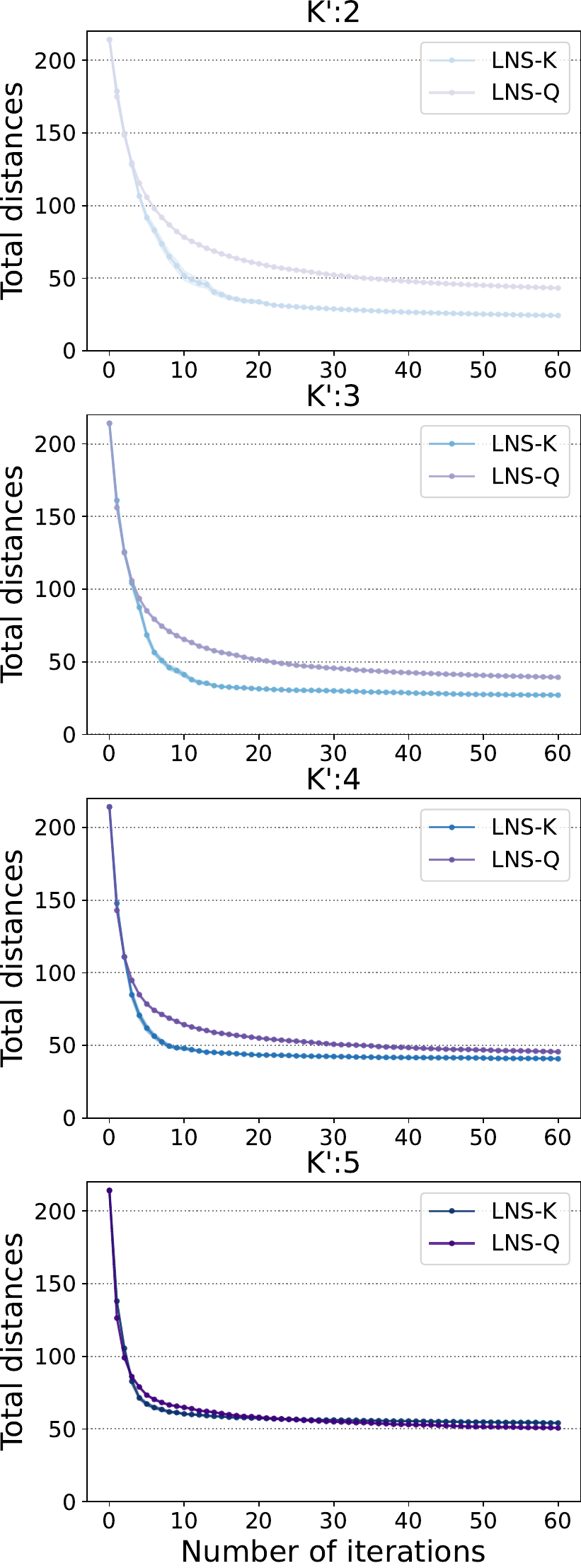}\\
        (a)
    \end{minipage}
    \begin{minipage}[t]{0.49\linewidth}
        \centering
        \includegraphics[width=\linewidth]{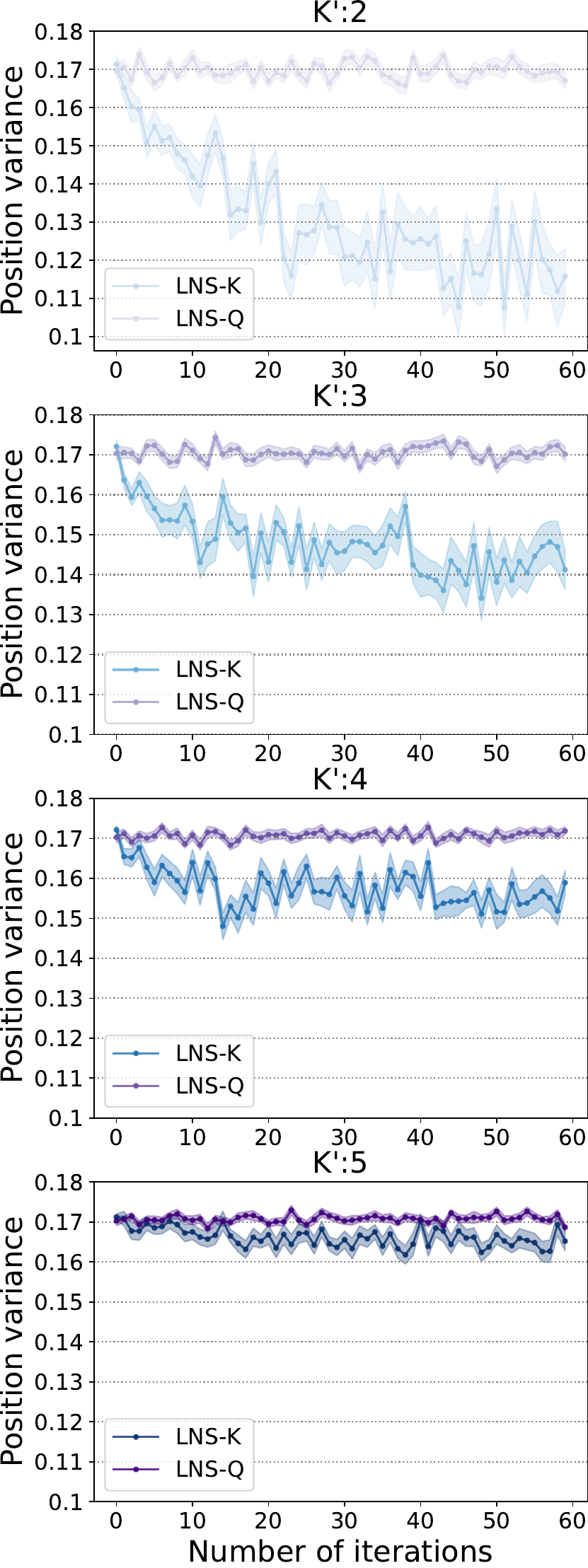}\\
        (b)
    \end{minipage}
    \caption{Iterative behavior of LNS-K and LNS-Q for subproblems with the same number of binary variables. The results are for a vehicle routing problem with 400 customers. Each row corresponds to $K' = 2, 3, 4$, and $5$ in LNS-K, and $n'$ in LNS-Q is chosen to match the corresponding subproblem size, as listed in Table~\ref{tab:matched_subproblem_sizes}. (a) Total distance of the current solution. LNS-K shows faster iterative improvement than LNS-Q, especially for smaller subproblems. (b) Position variance of the selected customer set. In LNS-K, the selected customer set becomes more spatially localized during the optimization, whereas the position variance in LNS-Q remains nearly constant.}    \label{fig:totaldistances_positionvariance_lns_k_and_q}
\end{figure}

We next compare the iterative improvement processes under matched subproblem sizes.
Figure~\ref{fig:totaldistances_positionvariance_lns_k_and_q}~(a) shows the total distance of the current solution as a function of the iteration number.
The pairs of $K'$ in LNS-K and $n'$ in LNS-Q are chosen so that the two methods have the same number of binary variables in the subproblem, as listed in Table~\ref{tab:matched_subproblem_sizes}.
The four rows correspond to $K'=2$, $3$, $4$, and $5$ in LNS-K.

For all four matched-size comparisons, the two methods start from the same total distance because they share the same initial solutions, and both methods reduce the total distance during the iteration process.
The subsequent trajectories are different between LNS-K and LNS-Q even though the numbers of binary variables in the subproblems are matched.
For $K'=2$ and $K'=3$, LNS-K decreases the total distance more rapidly than LNS-Q and reaches a shorter final total distance.
For $K'=4$, LNS-K also stays below LNS-Q after the early stage of the iteration.
For $K'=5$, the two curves become closer than those for smaller subproblems.

The difference between LNS-K and LNS-Q is therefore more visible for smaller matched subproblems.
This result shows that the effect of the subproblem construction rule appears not only in the final value but also in the iterative improvement process.

\subsection{Geometric evolution of selected customer sets}
\label{subsec:geometric_evolution}

We finally examine the geometric properties of the customer sets selected for the subproblems.
Figure~\ref{fig:totaldistances_positionvariance_lns_k_and_q}~(b) shows the position variance of the selected customer set as a function of the iteration number.
The position variance is defined as the sum of the variances of the two coordinates of the customers included in the subproblem.
A smaller value of this quantity corresponds to a more spatially localized customer set.

The position variance behaves differently in LNS-K and LNS-Q.
In LNS-Q, the position variance remains nearly constant during the iterations for all matched subproblem sizes.
In LNS-K, the position variance decreases during the iterations for the smaller subproblems.
The decrease is most visible for $K'=2$ and $K'=3$.
For $K'=4$, the position variance of LNS-K remains below that of LNS-Q after the early stage of the iteration.
For $K'=5$, the difference between the two methods is smaller.

These observations show that LNS-K and LNS-Q generate different geometric evolutions of the selected customer sets under the same number of binary variables.
Together with Fig.~\ref{fig:totaldistances_positionvariance_lns_k_and_q}~(a), Fig.~\ref{fig:totaldistances_positionvariance_lns_k_and_q}~(b) shows that the difference between the two construction rules appears both in the objective-function trajectory and in the spatial spread of the selected customer set.

\section{Discussion}
\label{sec:discussion}

This section discusses the implications of the results.
Section~\ref{subsec:size_not_sufficient} argues that the subproblem size alone is not a sufficient descriptor of a subproblem.
Section~\ref{subsec:geometric_interpretation} interprets the geometric localization observed for LNS-K.
Section~\ref{subsec:route_vs_qubo} discusses the different roles of the route structure and the QUBO structure.
Section~\ref{subsec:cross_platform} summarizes the cross-platform observations, and Section~\ref{subsec:limitations} states the limitations of the study.

\subsection{Subproblem size is not a sufficient descriptor}
\label{subsec:size_not_sufficient}

The main observation in Secs.~\ref{subsec:dependence_subproblem_construction} and \ref{subsec:iterative_improvement} is that the number of binary variables in a subproblem does not fully characterize the behavior of sequential optimization.
The number of binary variables is an important parameter because it directly controls the size of the QUBO problem submitted to an Ising machine.
However, two subproblems with the same number of binary variables can have different internal structures.
They can contain different sets of variables, different constraint relations, and different effective interactions with the fixed variables outside the subproblem.

This point is directly reflected in the comparison between LNS-K and LNS-Q.
In Fig.~\ref{fig:totaldistances_lns_k_and_q}, LNS-Q provides finer control of the subproblem size through $n'$ than LNS-K does through $K'$.
Even with this finer size control, the final total distance depends on the construction rule of the subproblem.
In Fig.~\ref{fig:totaldistances_positionvariance_lns_k_and_q}(a), the difference remains when the number of binary variables is matched between the two methods.
These results indicate that the subproblem size alone is not a sufficient descriptor of the subproblem used in Ising-machine-assisted sequential optimization.

The difference should not be interpreted as a statement that any specific structural quantity uniquely determines the performance.
The present results do not evaluate, for example, the graph density of the subproblem, the number of boundary couplings, or the distribution of effective fields induced by fixed variables.
The supported conclusion is more limited.
The way in which the variables are grouped into a subproblem affects the optimization trajectory even when the number of binary variables is fixed.

This observation is relevant for designing LNS with Ising machines.
A natural strategy is to tune the subproblem size so that the Ising machine can solve the subproblem with sufficient accuracy.
The present results show that this size tuning is necessary but not sufficient.
The variable-selection rule must also be considered as part of the subproblem design.

\subsection{Interpretation of the geometric localization}
\label{subsec:geometric_interpretation}

Figure~\ref{fig:totaldistances_positionvariance_lns_k_and_q}(b) shows that the customer sets selected by LNS-K become more spatially localized during the iterations, whereas those selected by LNS-Q remain nearly unchanged in their position variance.
This behavior is consistent with the way in which LNS-K constructs its subproblems.
LNS-K selects vehicles from the current solution and then includes the customers visited by those vehicles.
Thus, the selected customer set inherits the route structure of the current solution.

In a VRP, a good route often connects customers that are close to each other in space.
As the current solution is improved, the routes can become more spatially organized.
If the subproblem is generated from such routes, the selected customer set can also become more localized.
The decrease in the position variance in LNS-K may therefore reflect the spatial organization of the current routes during sequential optimization.

This interpretation is also compatible with the difference in the objective-function trajectories in Fig.~\ref{fig:totaldistances_positionvariance_lns_k_and_q}(a).
A spatially localized subproblem may contain variables related to local rearrangements of nearby routes.
Such a subproblem can be suitable for improving the current route structure once a moderately organized solution has been obtained.
This provides a possible explanation for why the route-structure-based construction gives larger iterative improvements than the QUBO-variable-based construction in the tested cases.

However, this interpretation remains qualitative.
The present study does not establish a causal relation between geometric localization and the improvement in the total distance.
We did not compute the correlation between the position variance and the improvement obtained in each iteration.
We also did not compare subproblems with controlled position variance.
Therefore, the decrease in the position variance should be regarded as an observed geometric feature of LNS-K, not as a proven cause of its performance.
Because the initial solutions are generated independently of the customer coordinates, this localization is not inherited from the initial solution and develops during the sequential optimization.

\subsection{Different roles of route structure and QUBO structure}
\label{subsec:route_vs_qubo}

The comparison between LNS-K and LNS-Q should not be understood as a comparison between a structured method and an unstructured random method.
LNS-Q uses the QUBO representation and the associated constraint relations to construct a subproblem.
It also allows the subproblem size to be varied more finely through $n'$.
This is an advantage when one wants to adjust the input size to the solving capability of an Ising machine.

LNS-K uses a different type of structure.
It selects vehicles and the customers visited by those vehicles in the current solution.
Thus, its subproblem is constructed from the route structure in the original VRP.
This construction is less generic than LNS-Q because it uses problem-domain information.
At the same time, it can directly preserve a semantic unit of the current solution.

The present results show that this semantic unit is relevant in the tested VRP.
Even when the number of binary variables is matched, LNS-K and LNS-Q show different iterative behaviors.
This indicates that preserving the route structure of the current solution can affect the optimization process.

The implication is not that a problem-specific construction is always superior to a QUBO-variable-based construction.
LNS-Q remains useful as a representation-level method that can be applied without explicitly defining route-based units.
The implication is that subproblem design should consider both the size of the subproblem and the structure inherited from the original problem or from the current solution.
For problems in which meaningful units can be identified, incorporating such units into the variable-selection rule can be an effective design principle.

\subsection{Cross-platform observations}
\label{subsec:cross_platform}

The results obtained with the two Ising machines have different roles in this study.
The D-Wave Advantage 4.1 results support the use of sequential subproblem optimization for small instances.
The Fixstars Amplify AE results are used for larger instances and for the detailed comparison between LNS-K and LNS-Q.
Therefore, the two sets of results should not be interpreted as a direct performance comparison between the two machines.

The common observation is that sequential subproblem optimization gives better results than the naive approach under the equal annealing-time condition used in this study.
This supports the basic motivation of using LNS with Ising machines.
By solving smaller subproblems repeatedly, the algorithm can improve a current solution without submitting the full QUBO problem at every iteration.

The detailed structure dependence, however, is based on the large-scale experiments with Fixstars Amplify AE.
In particular, the matched-size comparison between LNS-K and LNS-Q and the analysis of the position variance are not repeated on both hardware platforms.
Thus, the present results do not demonstrate that the subproblem-structure dependence is independent of the hardware implementation.
They show that, in the tested large-scale setting, the construction rule of the subproblem affects both the objective-function trajectory and the geometric evolution of the selected customer set.

This distinction is important for interpreting the scope of the results.
The cross-platform part supports the usefulness of sequential optimization as a general framework.
The structure-comparison part supports the importance of the subproblem construction rule in the tested VRP.
The present study does not claim hardware superiority, quantum advantage, or hardware-independent universality of the observed structure effect.

\subsection{Limitations}
\label{subsec:limitations}

The conclusions of this study have several limitations.
First, the detailed structure analysis was performed for VRPs with a specific QUBO formulation.
The observed difference between LNS-K and LNS-Q may depend on this formulation, on the penalty terms, and on the way in which feasible current solutions are updated.
It is therefore necessary to test other formulations and other constrained combinatorial optimization problems before making a more general claim.

Second, the matched-size comparison between LNS-K and LNS-Q is mainly based on the 400-customer experiments with Fixstars Amplify AE.
The results show a clear difference under the tested conditions, but they do not by themselves determine how the difference scales with the number of customers, the number of vehicles, or the distribution of customer locations.
A broader benchmark set would be needed to evaluate the robustness of the observed behavior.
In addition, a single customer configuration was used for each value of $N$, so the variability among the 30 runs reflects the random initial solutions, the random subproblem selection, and the stochastic behavior of the Ising machines rather than the variation among problem instances.

Third, the geometric analysis is descriptive.
The position variance was used to characterize the spatial spread of the selected customer set.
The decrease in the position variance in LNS-K was observed together with better iterative improvement.
However, the present study did not perform a direct correlation analysis between the position variance and the objective-function improvement.
It also did not construct artificial subproblems with controlled spatial spread.
Thus, the causal relation between spatial localization and performance remains open.

Fourth, the comparison between the naive approach and LNS was made under an equal annealing-time condition.
This condition is useful for comparing the time spent by the Ising machine.
However, it does not include all components of the end-to-end runtime, such as QUBO construction, communication, embedding, decoding, and postprocessing.
Therefore, the results should be interpreted as comparisons under a controlled annealing-time budget, not as complete wall-clock-time comparisons.

Fifth, the present work does not introduce new comparisons with classical optimization baselines beyond the existing numerical results.
The goal of this study is to clarify the role of the subproblem construction rule within Ising-machine-assisted sequential optimization.
A full evaluation against state-of-the-art classical vehicle routing heuristics is outside the scope of the present analysis.

Another limitation is the coarse control of the subproblem size in LNS-K.
In the present vehicle-based construction, the parameter $K'$ changes the number of selected vehicles.
The subproblem size therefore changes discretely and relatively coarsely.
In the tested range, the shortest final total distance in LNS-K is obtained at $K'=2$, which is the lower end of the examined values.
This observation suggests that finer control of the subproblem size may be useful while retaining the feasibility-preserving property of the route-based construction.
Developing such extensions is beyond the scope of the present paper.

These limitations do not change the main conclusion supported by Figs.~\ref{fig:totaldistances_lns_k_and_q} and \ref{fig:totaldistances_positionvariance_lns_k_and_q}.
Under matched subproblem sizes, the route-structure-based construction and the QUBO-variable-based construction show different optimization trajectories.
The selected customer sets also exhibit different geometric evolutions.
These observations indicate that the number of binary variables is not sufficient to characterize a subproblem and that the structural information inherited from the current solution should be considered in subproblem design.
\section{Summary and Conclusion}
\label{sec:summary_conclusion}

We investigated Ising-machine-assisted sequential optimization for VRPs.
The focus of this study was the role of subproblem construction in LNS.
We compared two construction rules.
LNS-K constructs a subproblem from selected vehicles and the customers visited by those vehicles in the current solution.
LNS-Q constructs a subproblem from QUBO variables and the associated constraint relations.
The central question was whether the construction rule affects the optimization behavior even when the number of binary variables in the subproblem is controlled.

First, we compared LNS with the naive approach under matched total annealing-time budgets.
In the naive approach, the full QUBO problem was submitted to an Ising machine once.
In LNS, smaller subproblems were solved sequentially, and the current solution was updated during the optimization process.
For the tested instances, LNS gave shorter total distances than the naive approach under the matched annealing-time condition.
This result supports the use of sequential subproblem optimization as a practical way to apply Ising machines to constrained VRP instances.
The comparison was made in terms of total annealing time and not in terms of end-to-end wall-clock time.

Second, we compared LNS-K and LNS-Q under matched subproblem sizes.
The number of binary variables in the subproblem was matched by choosing corresponding values of \(K'\) in LNS-K and \(n'\) in LNS-Q.
Even under this matched-size condition, the two methods showed different final total distances and different iterative trajectories.
In the tested 400-customer instance, LNS-K gave shorter final total distances than LNS-Q.
This result shows that the number of binary variables alone does not fully characterize the subproblem used in sequential optimization.

Third, we analyzed the geometric evolution of the customer sets selected for the subproblems.
In LNS-K, the position variance of the selected customer set decreased during the iterations.
This behavior indicates that the selected customer set became more spatially localized as the current solution was improved.
In contrast, the position variance in LNS-Q remained nearly constant in the matched-size comparisons.
This observation shows that the two construction rules differ not only in their objective-function trajectories but also in the geometric evolution of the selected customer sets.

These conclusions are limited to the tested VRP instances, the QUBO formulation used in this study, and the examined subproblem construction rules.
The present results do not establish a causal relation between spatial localization and the improvement in the total distance.
They nevertheless show that subproblem design cannot be reduced to size control alone.
Overall, subproblem construction, rather than the number of binary variables alone, shapes the optimization behavior, and the structure inherited from the current solution should therefore be treated as an integral part of subproblem design for Ising machines.

\section*{Acknowledgments}
The authors are grateful to Koshiro Fujimoto for valuable discussions.
This work was partially supported by the Japan Society for the Promotion of Science (JSPS) KAKENHI (Grant Number JP23H05447), the Council for Science, Technology, and Innovation (CSTI) through the Cross-ministerial Strategic Innovation Promotion Program (SIP), ``Promoting the application of advanced quantum technology platforms to social issues'' (Funding agency: QST), Japan Science and Technology Agency (JST) (Grant Number JPMJPF2221). The computations in this work were partially performed using the facilities of the Supercomputer Center, the Institute for Solid State Physics, The University of Tokyo. S. Tanaka wishes to express their gratitude to the World Premier International Research Center Initiative (WPI), MEXT, Japan, for their support of the Human Biology-Microbiome-Quantum Research Center (Bio2Q).

\bibliographystyle{jpsj}
\bibliography{reference}

\end{document}